\documentclass[%
 superscriptaddress,
 reprint,
 showpacs,preprintnumbers,
 nofootinbib,
 amsmath,amssymb,
 aps,
 prl,
 floatfix,
]{revtex4-1}

\usepackage{graphicx}% Include figure files

\usepackage{xcolor}

\usepackage{lipsum}

\usepackage{amssymb,amstext,amsthm,amsfonts,amsfonts}
\usepackage[utf8]{inputenc}

\usepackage[linesnumbered,ruled]{algorithm2e}
\usepackage{url}
\usepackage{soul}
\usepackage{bm}
\usepackage[normalem]{ulem}
\usepackage{multirow, makecell}

\makeatletter
\def\@fnsymbol#1{\ensuremath{\ifcase#1\or
   *\or \dagger\or \ddagger\or \mathsection\or \mathparagraph\or \|\or
   **\or \dagger\dagger\or \ddagger\ddagger\or
   \mathsection\mathsection\or \mathparagraph\mathparagraph\or \|\|\or
   ***\or \dagger\dagger\dagger\or \ddagger\ddagger\ddagger\or
   \mathsection\mathsection\mathsection\or \mathparagraph\mathparagraph\mathparagraph\or \|\|\|\or
   ****\or \dagger\dagger\dagger\dagger\or \ddagger\ddagger\ddagger\ddagger
   \else\@ctrerr\fi}}
\makeatother

\usepackage{hyperref}
\hypersetup{
    hidelinks,               % Removes all borders/boxes
    colorlinks=true,         % Enables colored links
    citecolor=teal,          % Color for citations
    linkcolor=teal,           % Color for internal links
    urlcolor=teal           % External URLs
}

\begin{document}
    \title{First differential measurement of the single $\mathbf{\pi}^+$ production cross section in neutrino neutral-current scattering}

%%%%%%%%%%%%%%%%%%%%%%%%%%%%%%%%%%%%%%%%%%%%%%%%%%%%%%%%%%%%%%
% T2K author list generated on Fri Jun 13 16:31:18 2025
% setting: extra = False
%         author list from archive (starting October 20 2024 until March 06 2025)
%         exemption(s) granted to: ashida,rgonzjim,jgrosa,gdmegias,jac,jgm,hmatthias,njachowicz,akumar,,edatkin,martind,kniewczas,scap,vedantha,fcadoux,fsanchez,evila,amunoz,nova,pdunne,dlast,adrienblanchet,mkabirne,barker,morgan_wascko,bdl,vlglagolev,slava,mtzanov,kutter,lmachado,litchfieldp,egoodman,yyang,kiseevavi,akaboth,kmahtani,dseppala,acraplet,alex,kudenko,oleg,erofeev,kurochka,matveev,shaykhiev,greina,asrivast,nicklatham,nakahata,rubbiaa,laveder,mezzetto,Forza,mgrassi,damartin,denhaff,wilking
% Number of authors = 382
%%%%%%%%%%%%%%%%%%%%%%%%%%%%%%%%%%%%%%%%%%%%%%%%%%%%%%%%%%%%%%

\newcommand{\INSTHD}{\affiliation{University Autonoma Madrid, Department of Theoretical Physics, 28049 Madrid, Spain}}
\newcommand{\INSTFE}{\affiliation{Boston University, Department of Physics, Boston, Massachusetts, U.S.A.}}
\newcommand{\INSTD}{\affiliation{University of British Columbia, Department of Physics and Astronomy, Vancouver, British Columbia, Canada}}
\newcommand{\INSTGA}{\affiliation{University of California, Irvine, Department of Physics and Astronomy, Irvine, California, U.S.A.}}
\newcommand{\INSTI}{\affiliation{IRFU, CEA, Universit\'e Paris-Saclay, F-91191 Gif-sur-Yvette, France}}
\newcommand{\INSTGB}{\affiliation{University of Colorado at Boulder, Department of Physics, Boulder, Colorado, U.S.A.}}
\newcommand{\INSTFH}{\affiliation{Duke University, Department of Physics, Durham, North Carolina, U.S.A.}}
\newcommand{\INSTJA}{\affiliation{E\"{o}tv\"{o}s Lor\'{a}nd University, Department of Atomic Physics, Budapest, Hungary}}
\newcommand{\INSTEF}{\affiliation{ETH Zurich, Institute for Particle Physics and Astrophysics, Zurich, Switzerland}}
\newcommand{\INSTIG}{\affiliation{VNU University of Science, Vietnam National University, Hanoi, Vietnam}}
\newcommand{\INSTIE}{\affiliation{CERN European Organization for Nuclear Research, CH-1211 Gen\'eve 23, Switzerland}}
\newcommand{\INSTEG}{\affiliation{University of Geneva, Section de Physique, DPNC, Geneva, Switzerland}}
\newcommand{\INSTHJ}{\affiliation{University of Glasgow, School of Physics and Astronomy, Glasgow, United Kingdom}}
\newcommand{\INSTJG}{\affiliation{Ghent University, Department of Physics and Astronomy, Proeftuinstraat 86, B-9000 Gent, Belgium}}
\newcommand{\INSTDG}{\affiliation{H. Niewodniczanski Institute of Nuclear Physics PAN, Cracow, Poland}}
\newcommand{\INSTCB}{\affiliation{High Energy Accelerator Research Organization (KEK), Tsukuba, Ibaraki, Japan}}
\newcommand{\INSTIB}{\affiliation{University of Houston, Department of Physics, Houston, Texas, U.S.A.}}
\newcommand{\INSTED}{\affiliation{Institut de Fisica d'Altes Energies (IFAE) - The Barcelona Institute of Science and Technology, Campus UAB, Bellaterra (Barcelona) Spain}}
\newcommand{\INSTJC}{\affiliation{Institut f\"ur Physik, Johannes Gutenberg-Universit\"at Mainz, Staudingerweg 7, 55128 Mainz, Germany}}
\newcommand{\INSTHH}{\affiliation{Institute For Interdisciplinary Research in Science and Education (IFIRSE), ICISE, Quy Nhon, Vietnam}}
\newcommand{\INSTEI}{\affiliation{Imperial College London, Department of Physics, London, United Kingdom}}
\newcommand{\INSTGF}{\affiliation{INFN Sezione di Bari and Universit\`a e Politecnico di Bari, Dipartimento Interuniversitario di Fisica, Bari, Italy}}
\newcommand{\INSTBE}{\affiliation{INFN Sezione di Napoli and Universit\`a di Napoli, Dipartimento di Fisica, Napoli, Italy}}
\newcommand{\INSTBF}{\affiliation{INFN Sezione di Padova and Universit\`a di Padova, Dipartimento di Fisica, Padova, Italy}}
\newcommand{\INSTBD}{\affiliation{INFN Sezione di Roma and Universit\`a di Roma ``La Sapienza'', Roma, Italy}}
\newcommand{\INSTEB}{\affiliation{Institute for Nuclear Research of the Russian Academy of Sciences, Moscow, Russia}}
\newcommand{\INSTHI}{\affiliation{International Centre of Physics, Institute of Physics (IOP), Vietnam Academy of Science and Technology (VAST), 10 Dao Tan, Ba Dinh, Hanoi, Vietnam}}
\newcommand{\INSTJD}{\affiliation{ILANCE, CNRS – University of Tokyo International Research Laboratory, Kashiwa, Chiba 277-8582, Japan}}
\newcommand{\INSTHA}{\affiliation{Kavli Institute for the Physics and Mathematics of the Universe (WPI), The University of Tokyo Institutes for Advanced Study, University of Tokyo, Kashiwa, Chiba, Japan}}
\newcommand{\INSTID}{\affiliation{Keio University, Department of Physics, Kanagawa, Japan}}
\newcommand{\INSTIF}{\affiliation{King's College London, Department of Physics, Strand, London WC2R 2LS, United Kingdom}}
\newcommand{\INSTCC}{\affiliation{Kobe University, Kobe, Japan}}
\newcommand{\INSTCD}{\affiliation{Kyoto University, Department of Physics, Kyoto, Japan}}
\newcommand{\INSTEJ}{\affiliation{Lancaster University, Physics Department, Lancaster, United Kingdom}}
\newcommand{\INSTII}{\affiliation{Lawrence Berkeley National Laboratory, Berkeley, California, U.S.A.}}
\newcommand{\INSTBA}{\affiliation{Ecole Polytechnique, IN2P3-CNRS, Laboratoire Leprince-Ringuet, Palaiseau, France}}
\newcommand{\INSTFC}{\affiliation{University of Liverpool, Department of Physics, Liverpool, United Kingdom}}
\newcommand{\INSTFI}{\affiliation{Louisiana State University, Department of Physics and Astronomy, Baton Rouge, Louisiana, U.S.A.}}
\newcommand{\INSTIH}{\affiliation{Joint Institute for Nuclear Research, Dubna, Moscow Region, Russia}}
\newcommand{\INSTHB}{\affiliation{Michigan State University, Department of Physics and Astronomy,  East Lansing, Michigan, U.S.A.}}
\newcommand{\INSTCE}{\affiliation{Miyagi University of Education, Department of Physics, Sendai, Japan}}
\newcommand{\INSTDF}{\affiliation{National Centre for Nuclear Research, Warsaw, Poland}}
\newcommand{\INSTFJ}{\affiliation{State University of New York at Stony Brook, Department of Physics and Astronomy, Stony Brook, New York, U.S.A.}}
\newcommand{\INSTEH}{\affiliation{STFC, Rutherford Appleton Laboratory, Harwell Oxford,  and  Daresbury Laboratory, Warrington, United Kingdom}}
\newcommand{\INSTGJ}{\affiliation{Okayama University, Department of Physics, Okayama, Japan}}
\newcommand{\INSTCF}{\affiliation{Osaka Metropolitan University, Department of Physics, Osaka, Japan}}
\newcommand{\INSTGG}{\affiliation{Oxford University, Department of Physics, Oxford, United Kingdom}}
\newcommand{\INSTIC}{\affiliation{University of Pennsylvania, Department of Physics and Astronomy,  Philadelphia, Pennsylvania, U.S.A.}}
\newcommand{\INSTGC}{\affiliation{University of Pittsburgh, Department of Physics and Astronomy, Pittsburgh, Pennsylvania, U.S.A.}}
\newcommand{\INSTGD}{\affiliation{University of Rochester, Department of Physics and Astronomy, Rochester, New York, U.S.A.}}
\newcommand{\INSTHC}{\affiliation{Royal Holloway University of London, Department of Physics, Egham, Surrey, United Kingdom}}
\newcommand{\INSTBC}{\affiliation{RWTH Aachen University, III. Physikalisches Institut, Aachen, Germany}}
\newcommand{\INSTJF}{\affiliation{School of Physics and Astronomy, University of Minnesota, Minneapolis, Minnesota, U.S.A.}}
\newcommand{\INSTJB}{\affiliation{Departamento de F\'isica At\'omica, Molecular y Nuclear, Universidad de Sevilla, 41080 Sevilla, Spain}}
\newcommand{\INSTFB}{\affiliation{University of Sheffield, School of Mathematical and Physical Sciences, Sheffield, United Kingdom}}
\newcommand{\INSTDI}{\affiliation{University of Silesia, Institute of Physics, Katowice, Poland}}
\newcommand{\INSTIA}{\affiliation{SLAC National Accelerator Laboratory, Stanford University, Menlo Park, California, U.S.A.}}
\newcommand{\INSTBB}{\affiliation{Sorbonne Universit\'e, CNRS/IN2P3, Laboratoire de Physique Nucl\'eaire et de Hautes Energies (LPNHE), Paris, France}}
\newcommand{\INSTJE}{\affiliation{South Dakota School of Mines and Technology, 501 East Saint Joseph Street, Rapid City, SD 57701, United States}}
\newcommand{\INSTCH}{\affiliation{University of Tokyo, Department of Physics, Tokyo, Japan}}
\newcommand{\INSTBJ}{\affiliation{University of Tokyo, Institute for Cosmic Ray Research, Kamioka Observatory, Kamioka, Japan}}
\newcommand{\INSTCG}{\affiliation{University of Tokyo, Institute for Cosmic Ray Research, Research Center for Cosmic Neutrinos, Kashiwa, Japan}}
\newcommand{\INSTHF}{\affiliation{Institute of Science Tokyo, Department of Physics, Tokyo}}
\newcommand{\INSTGI}{\affiliation{Tokyo Metropolitan University, Department of Physics, Tokyo, Japan}}
\newcommand{\INSTHG}{\affiliation{Tokyo University of Science, Faculty of Science and Technology, Department of Physics, Noda, Chiba, Japan}}
\newcommand{\INSTB}{\affiliation{TRIUMF, Vancouver, British Columbia, Canada}}
\newcommand{\INSTJH}{\affiliation{University of Toyama, Department of Physics, Toyama, Japan}}
\newcommand{\INSTDJ}{\affiliation{University of Warsaw, Faculty of Physics, Warsaw, Poland}}
\newcommand{\INSTDH}{\affiliation{Warsaw University of Technology, Institute of Radioelectronics and Multimedia Technology, Warsaw, Poland}}
\newcommand{\INSTIJ}{\affiliation{Tohoku University, Faculty of Science, Department of Physics, Miyagi, Japan}}
\newcommand{\INSTFD}{\affiliation{University of Warwick, Department of Physics, Coventry, United Kingdom}}
\newcommand{\INSTEA}{\affiliation{Wroclaw University, Faculty of Physics and Astronomy, Wroclaw, Poland}}
\newcommand{\INSTHE}{\affiliation{Yokohama National University, Department of Physics, Yokohama, Japan}}
\newcommand{\INSTH}{\affiliation{York University, Department of Physics and Astronomy, Toronto, Ontario, Canada}}

\INSTHD
\INSTFE
\INSTD
\INSTGA
\INSTI
\INSTGB
\INSTFH
\INSTJA
\INSTEF
\INSTIG
\INSTIE
\INSTEG
\INSTHJ
\INSTJG
\INSTDG
\INSTCB
\INSTIB
\INSTED
\INSTJC
\INSTHH
\INSTEI
\INSTGF
\INSTBE
\INSTBF
\INSTBD
\INSTEB
\INSTHI
\INSTJD
\INSTHA
\INSTID
\INSTIF
\INSTCC
\INSTCD
\INSTEJ
\INSTII
\INSTBA
\INSTFC
\INSTFI
\INSTIH
\INSTHB
\INSTCE
\INSTDF
\INSTFJ
\INSTEH
\INSTGJ
\INSTCF
\INSTGG
\INSTIC
\INSTGC
\INSTGD
\INSTHC
\INSTBC
\INSTJF
\INSTJB
\INSTFB
\INSTDI
\INSTIA
\INSTBB
\INSTJE
\INSTCH
\INSTBJ
\INSTCG
\INSTHF
\INSTGI
\INSTHG
\INSTB
\INSTJH
\INSTDJ
\INSTDH
\INSTIJ
\INSTFD
\INSTEA
\INSTHE
\INSTH

\author{K.\,Abe}\INSTBJ
\author{S.\,Abe}\INSTBJ
\author{R.\,Akutsu}\INSTCB
\author{H.\,Alarakia-Charles}\INSTEJ
\author{Y.I.\,Alj Hakim}\INSTFB
\author{S.\,Alonso Monsalve}\INSTEF
\author{L.\,Anthony}\INSTEI
\author{S.\,Aoki}\INSTCC
\author{K.A.\,Apte}\INSTEI
\author{T.\,Arai}\INSTCH
% ...

\collaboration{The T2K Collaboration}\noaffiliation

\begin{abstract}
\noindent
 Since its first observation in the 1970s, neutrino-induced neutral-current single positive pion production (NC1$\pi^+$) has remained an elusive and poorly understood interaction channel. This process is a significant background in neutrino oscillation experiments and studying it further is critical for the physics program of next-generation accelerator-based neutrino oscillation experiments. In this Letter we present the first double-differential cross-section measurement of NC1$\pi^+$ interactions using data from the ND280 detector of the T2K experiment collected in $\nu$-beam mode. The measured flux-averaged integrated cross-section is $ \sigma = (6.07 \pm 1.22 )\times 10^{-41} \,\, \text{cm}^2/\text{nucleon}$. We compare the results on a hydrocarbon target to the predictions of several neutrino interaction generators and final-state interaction models. While model predictions agree with the differential results, the data shows a weak preference for a cross-section normalization approximately 30\% higher than predicted by most models studied in this Letter.
 
\end{abstract}

\maketitle

{\em Introduction.} Neutrino physics plays a central role in modern high-energy physics. Following the discovery of neutrino oscillations~\cite{Super-Kamiokande:1998kpq, SNO:2002tuh} several experiments are ongoing and planned with the aim of precisely characterizing the phenomenology of neutrino oscillations. Neutrino mixing is characterized by three-flavor mixing, depending on two mass splittings ($\Delta m^2_{23}$, $\Delta m^2_{21}$) and described by the Pontecorvo-Maki-Nakagawa-Sakata (PMNS) matrix, consisting of three mixing angles ($\theta_{13}$, $\theta_{23}$, $\theta_{12}$) and one Charge-Parity (CP) violating phase ($\delta_{\mathrm{CP}})$. Several parameters are already known with a few percent precision~\cite{Esteban:2024eli}, but major unknowns remain: the sign of $\Delta m^2_{23}$, known as the neutrino mass ordering; whether CP symmetry is violated or not, which is primarily determined by the value of $\delta_{\mathrm{CP}}$; and, the octant of $\theta_{23}$, determining if $\theta_{23}$ is smaller than, larger than or equal to $45^\circ$.

For existing and planned neutrino beam experiments, the experimental sensitivity to the unknown neutrino oscillation parameters becomes maximal for neutrino energies ranging from hundreds of MeV to a few GeV. In this energy range, multiple neutrino interaction mechanisms are possible, and poorly understood nuclear effects contribute significant systematic uncertainties~\cite{NuSTEC:2017hzk, NuSTEC:2019lqd, NuSTEC:2020nsl, T2K:2023smv}. Over the past two decades, more than twenty measurements have investigated neutrino single $\pi^0$ production. Of these, more than ten have studied neutral-current (NC) interactions~\cite{MicroBooNE:2024sec, ArgoNeuT:2015ldo, K2K:2004qpv, MicroBooNE:2022zhr, MINERvA:2016uck, MiniBooNE:2009dxl, MiniBooNE:2008mmr, MINOS:2016yyz, NOMAD:2009idt, NOvA:2019bdw, SciBooNE:2009nlf, SciBooNE:2010lca, T2K:2017epu}, with the remainder studying charged-current (CC) interactions~\cite{MicroBooNE:2024bnl, K2K:2010xeb, MicroBooNE:2018neo, MINERvA:2020anu, MINERvA:2016sfc, MINERvA:2015slz, MINERvA:2017okh, MiniBooNE:2010cxl, NOvA:2023uxq}. Similarly, more than twenty analyses have examined single $\pi^\pm$ production, though exclusively through CC interactions~\cite{MINERvA:2022esg, ArgoNeuT:2018und, ArgoNeuT:2014uwh, K2K:2008tus, K2K:2005uiu, MINERvA:2022djk, MINERvA:2014ogb, MINERvA:2016sfc, MINERvA:2014ani, MINERvA:2019rhx, MINERvA:2017ipy, MiniBooNE:2010eis, MiniBooNE:2009koj, SciBooNE:2008bzb, T2K:2023xlh, T2K:2021naz, T2K:2016cbz, T2K:2016soz, T2K:2019yqu, NINJA:2022zbi, NOMAD:2001vdk}. However, measurements of neutrino NC single $\pi^\pm$ production (NC1$\pi^{\pm}$), primarily generated by $\nu+p \rightarrow \nu + n + \pi^+$ interactions, have been notably absent from the literature.

In this Letter, we address this long-standing absence of results for this channel by presenting the first differential cross-section measurement of NC1$\pi^+$ interactions, reported as a function of the pion momentum and angle.

\begin{figure}[ht!]
  \centering
  \includegraphics[width=0.4\textwidth]{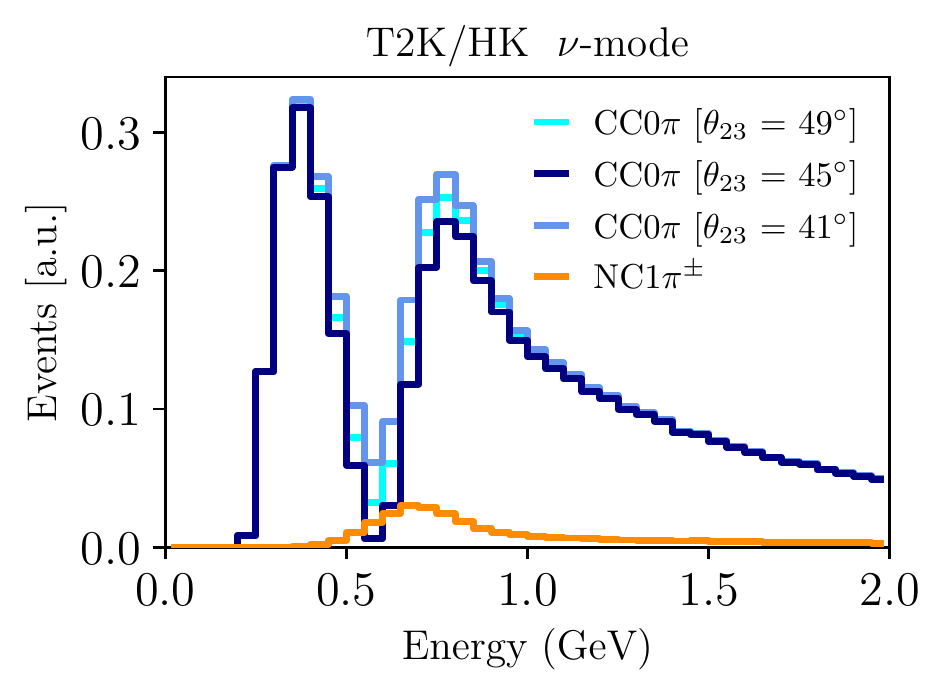}
  \includegraphics[width=0.4\textwidth]{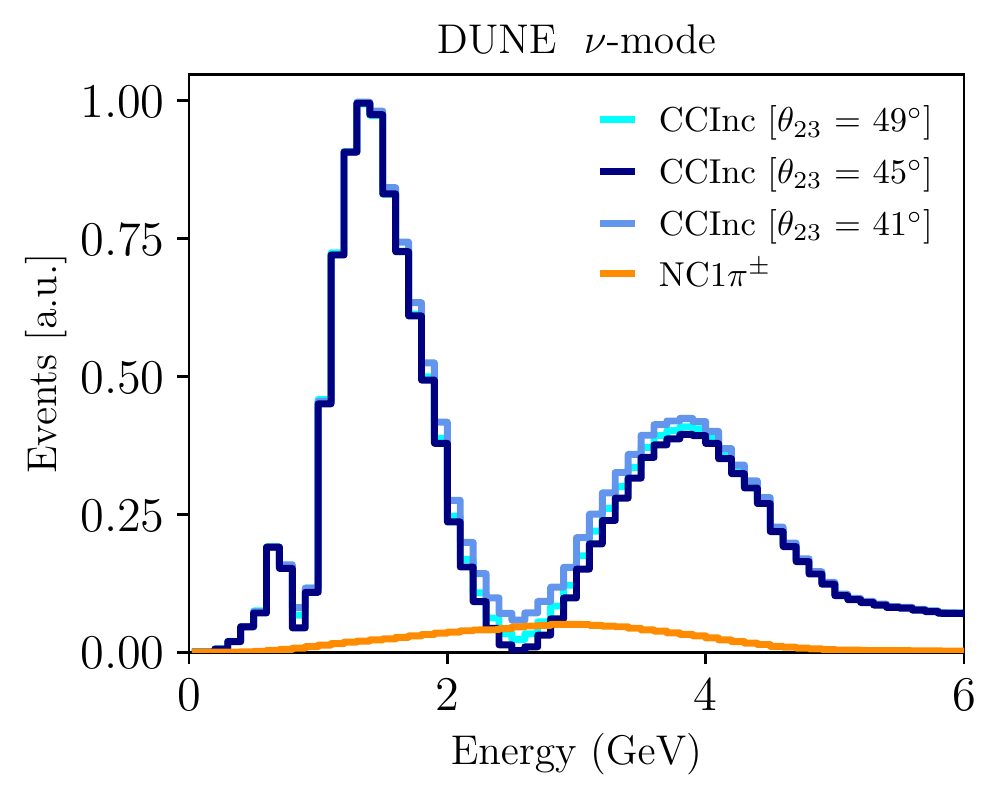}
  \caption{Expected event rate, in arbitrary units, for CC0$\pi$ including $\mu$ and $\pi$ Cherenkov thresholds, (T2K/Hyper-K) and CC-inclusive (DUNE) muon-neutrino interactions at the Hyper-K (top) and DUNE (bottom) experiments, shown as a function of true neutrino energy. Also overlaid is the expected rate of the NC1$\pi^{\pm}$ background in each case. The oscillation parameters used are from NuFit5.0~\cite{Esteban:2020cvm}, and use GENIEv3~\cite{Andreopoulos:2009rq,Andreopoulos:2015wxa} truth information to predict the interaction rates using T2K~\cite{T2K:2012bge} and DUNE~\cite{DUNE:2020ypp, DUNE:2020jqi} neutrino fluxes.}
  \label{fig:rate_comp}
\end{figure}

{\em Existing Measurements.} The Gargamelle~\cite{GargamelleNeutrinoPropane:1977hya} and ANL~\cite{Derrick:1980nr} bubble chamber experiments measured NC1$\pi^{\pm}$ interactions over four decades ago. At the time, their focus was demonstrating consistency with the newly proposed Standard Model of particle physics, confirming the existence of these interactions. One measurement of the NC1$\pi^{+}$ cross section exists in the scientific literature\footnote{To the best knowledge of the authors, details on the analysis methodology have been lost.}, a re-analysis of the Gargamelle reported event rates, collected at an average neutrino energy of 2~GeV~\cite{Zeller:2003ey}. 

{\em NC1$\pi^\pm$ in LBE.} Fig.~\ref{fig:rate_comp} shows the predicted oscillated muon-neutrino interaction rate as a function of neutrino energy for two future long-baseline neutrino experiments (LBE): Hyper-Kamiokande~\cite{Hyper-Kamiokande:2018ofw}, with an unoscillated neutrino energy peak at $\approx$~0.6 GeV and a water Cherenkov far detector, which focuses on CC0$\pi$ events; and DUNE~\cite{DUNE:2020ypp}, with a peak unoscillated neutrino energy of $\approx$~2.5 GeV and a liquid argon far detector, using all CC events. The neutrino energy peak is deliberately aligned near the energy where maximum disappearance of the muon flavor is expected for the experiment's baseline, creating a characteristic ``dip'' in the observed energy distribution. The depth of this dip and its position are used to determine the value of the $\theta_{23}$ and $\Delta m^2_{23}$ parameters. Overlaid is NC1$\pi^{\pm}$ predicted event rate, which is a hard to reject background as pions are frequently mis-identified as muons in both water Cherenkov detectors and liquid argon detectors. Due to oscillations, the CC event rate associated to muon-neutrinos is strongly reduced near the dip, however, all neutrino flavors contribute to the NC rate. As a result, NC1$\pi^{\pm}$ interactions can add a significant background contribution for muon-disappearance samples. In the T2K experiment, presented next in this Letter, statistical and systematic uncertainties for this channel are comparable. In the near future, Hyper-Kamiokande (HK) and DUNE, are expected to select hundreds of NC1$\pi^{\pm}$ events as part of their oscillation samples. Therefore, better understanding the NC1$\pi^{\pm}$ cross section and constraining its modeling uncertainty is important for the precision oscillation physics goals of next-generation experiments. 

{\em Experimental Setup. } The Tokai-to-Kamioka (T2K) long-baseline neutrino experiment~\cite{T2K:2011qtm} studies neutrino oscillations using a highly pure muon (anti)neutrino beam produced by the neutrino facility at the J-PARC proton accelerator. Data collection is divided in $\nu$-beam or ${\bar{\nu}}$-beam runs. The experiment combines measurements at two detector sites: the near detector facility, located 280 meters from the proton beam target, including the main near detector ND280, which was used to collect the data for this study; and the far detector facility, located 295 km downstream, is the Super-Kamiokande (SK) detector~\cite{Super-Kamiokande:2002weg}. Both ND280 and SK are placed 2.5$^\circ$ off-axis, resulting in a narrow neutrino energy spectrum peaking at 0.6~GeV~\cite{T2K:2012bge, E899:1995bzq}. 

The analysis utilizes ND280 data recorded in $\nu$-beam mode. ND280~\cite{T2K:2011qtm} is a magnetized particle detector made of several sub-detector modules, including: two Fine-Grained-Detectors (FGD1 and FGD2)~\cite{T2KND280FGD:2012umz}, three gaseous Time-Projection-Chambers (TPCs)~\cite{T2KND280TPC:2010nnd} and an Electromagnetic Calorimeter (ECal)~\cite{T2KUK:2013wkh}. Together, they provide rich information used to reconstruct the particle type, its trajectory and its momentum.

{\em Analysis samples.} Events with a vertex reconstructed in the fully-active hydrocarbon target FGD1 are classified in four different regions: a signal-enriched sample and three background-enriched samples. The signal definition includes all $\nu$ and $\bar{\nu}$ interactions without charged leptons in the final state, a single positive pion, any number of protons with momentum below the detection threshold of 200~MeV/c (See details in Ref.~\cite{Abe:2025qki}), and no other mesons or other charged particles in the final state. Due to ND280's acceptance and particle identification capabilities, the measurement is reported in the kinematic Region of Interest (RoI) defined by the conditions $\cos\theta_{\pi^{+}}>$0.5 and $0.2<p_{\pi^+}< 1.0$~GeV/$c$. The neutrino generator NEUT v5.4.0~\cite{Hayato:2009zz, Hayato:2021heg} is used with T2K's flux and detector simulation pipeline~\cite{T2K:2011qtm} to predict the rate of these events. Over 200 signal events are expected overall, with approximately 165 in the signal sample at 30.5\% efficiency and 51.4\% purity, and an integrated detector uncertainty for the predicted event rate below 5\%. Three analysis sidebands enriched in background events are used in the analysis. This includes a sample with 81.0\% purity in $\bar{\nu}_\mu$~CC events, the most numerous background events in the signal sample corresponding to 28.2\% of its selected events. The other sidebands are enriched in events with additional tracks in the neutrino target and events where the main track is identified as a proton instead of a $\pi^+$. Full selection details and performance metrics are presented in Ref.~\cite{Abe:2025qki}.

\begin{figure}[htpb!]
\centering
\includegraphics[width=0.49\textwidth]{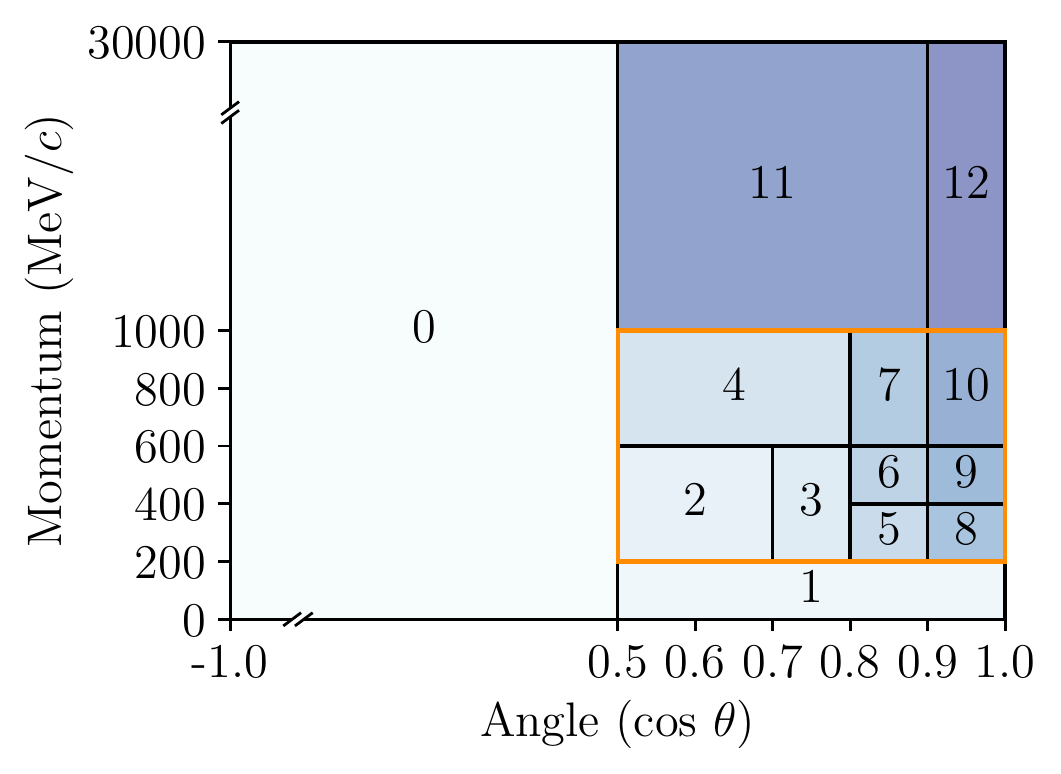}
\caption{Scheme of the true pion kinematic bins where the cross section is measured. Bin IDs are shown within their associated kinematic regions. The region of interest, were we report the measurement, is enclosed by an orange solid line and corresponds to bins 2-10. Color shades are used to aid visualization but have no physical meaning.}
\label{fig:bins_map}
\end{figure}

{\em Analysis strategy.} The NC1$\pi^+$ cross section is measured using an unregularized binned maximum likelihood fit, the same method described in earlier T2K measurements, e.g. in Refs.~\cite{T2K:2023qjb,T2K:2023xlh}. The procedure consists of numerically optimizing the parameters of interest, i.e. scaling factors for the number of signal events, together with nuisance parameters describing plausible variations of T2K's flux, detector and background cross-section models. The agreement with data is simultaneously maximized in the signal and background-enriched samples. This method provides a data-driven background and flux constraint while unfolding detector effects and providing a post-fit prediction of the expected signal events in every analysis bin, which is then used to calculate the signal cross section. In total, the flux-averaged cross section is unfolded in 13 bins in pion angle and momentum, nine of them contained in the RoI. The binning scheme is presented in Fig.~\ref{fig:bins_map}.
The model independence of the cross-section extraction was validated through extensive tests using alternative datasets with controlled variations. In all cases, small or negligible measurement biases were observed, confirming the robustness of the method.

{\em Interaction Models.} We compare the extracted double-differential and integrated cross section to several interaction generators using NUISANCE~\cite{Stowell:2016jfr}: NEUT v5.6.2~\cite{Hayato:2002sd,Hayato:2009,Hayato:2021heg}, NuWro v19.02.2~\cite{Golan:2012wx,Golan2012nuwro} and GENIEv3~\cite{Andreopoulos:2009rq,Andreopoulos:2015wxa} using the \texttt{AR23\_20i} configuration. These represent a range of commonly used models, which differ in a variety of details including hard scattering processes, nuclear modeling and re-interactions of interaction products within the nucleus (Final State Interactions, FSI). Additionally, we investigate the role of FSI using five different GENIE model variations, where the only change is the FSI model used. These represent a broad range of models of varying levels of sophistication ranging from no FSI, and full semi-classical cascade models. They include: the INTRANUKE~\cite{intranuke} hA single-step empirical model (\texttt{G18\_10a}); the INTRANUKE hN full intranuclear cascade (\texttt{G18\_10b}); GENIE's implementation of the INCL cascade~\cite{PhysRevC.91.034602,Cugnon:2016ghr} (\texttt{G18\_10c}); the GEANT4 Bertini cascade~\cite{Wright:2015xia} (\texttt{G18\_10d}); and an unphysical model in which all final-state interactions have been turned off (\texttt{G18\_10X}). \texttt{AR23\_20i} uses the hA model and is identical to \texttt{G18\_10a} for NC pion production, except that the free nucleon pion production tuning from Ref.~\cite{GENIE:2021zuu} is applied.

\begin{figure}[htpb!]
\centering
\includegraphics[width=0.49\textwidth]{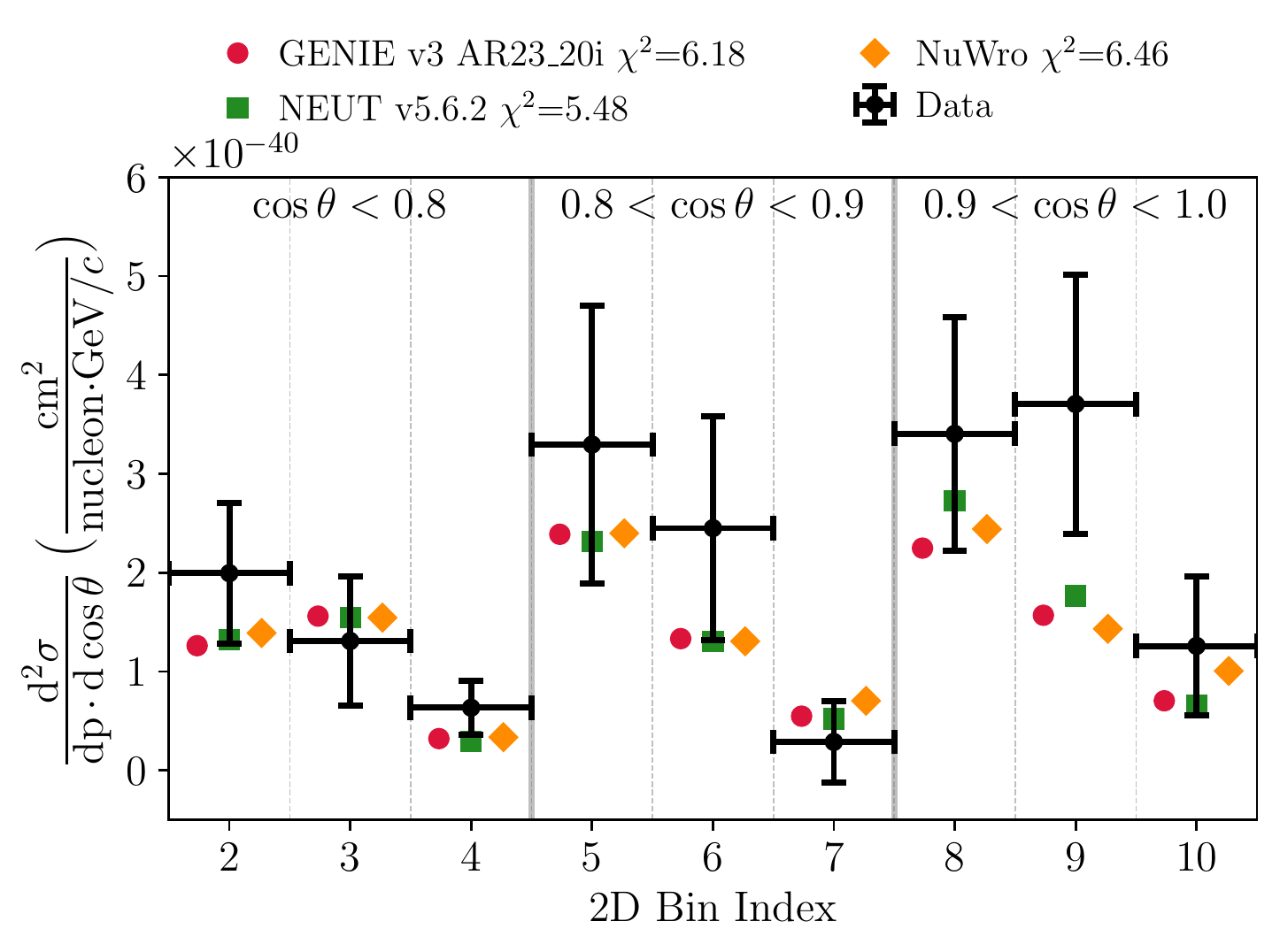}
\caption{Measured NC1$\pi^+$ double-differential cross section compared to generator predictions.}
\label{fig:result_vs_generators}
\end{figure}

\begin{figure}[htpb!]
\centering
\includegraphics[width=0.49\textwidth]{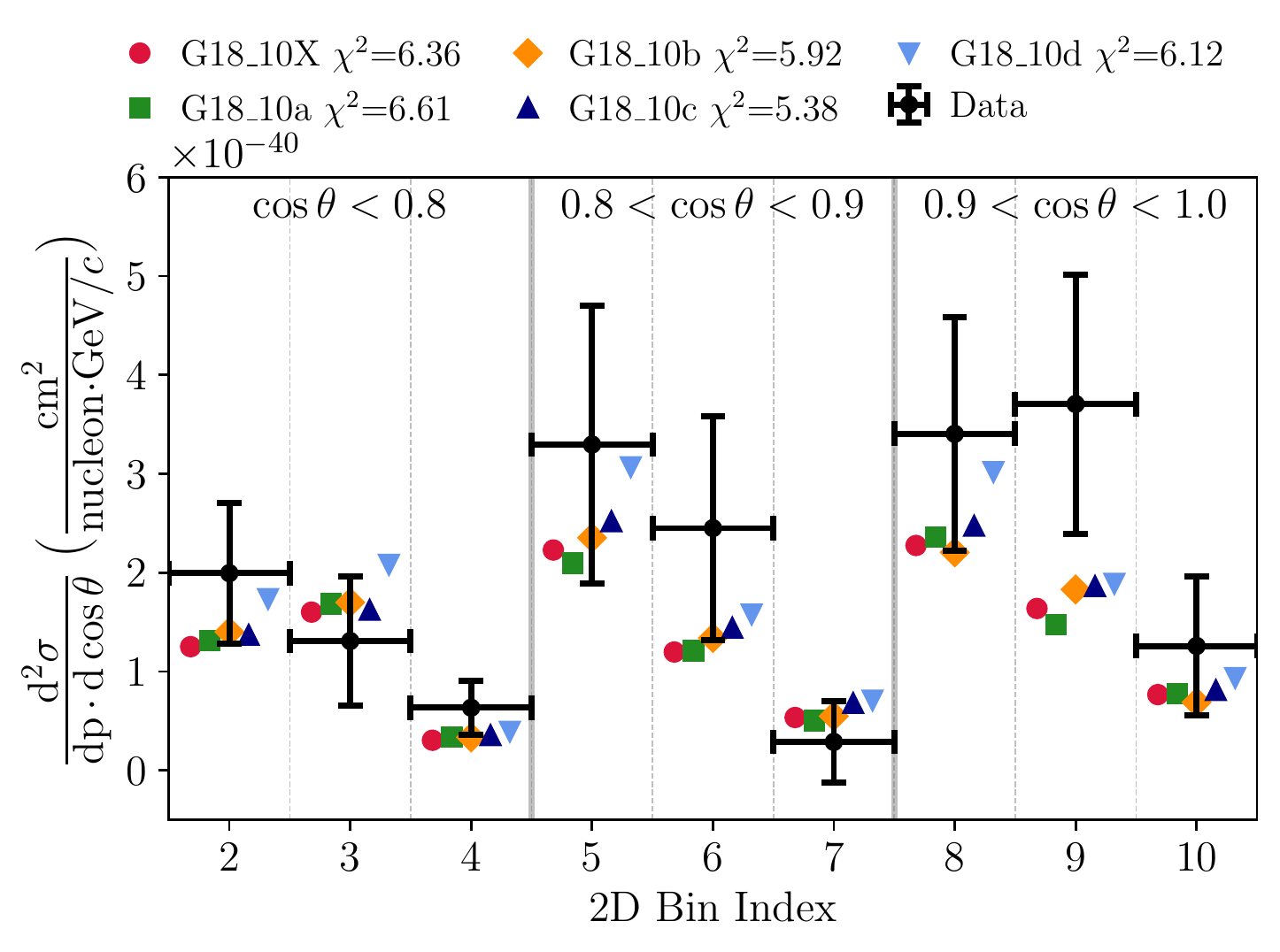}
\caption{Measured NC1$\pi^+$ double-differential cross section compared to various FSI predictions.}
\label{fig:result_vs_FSI}
\end{figure}

{\em Results.} The measured data cross section is compared to different neutrino interaction generator predictions in Fig.~\ref{fig:result_vs_generators} and FSI models in Fig.~\ref{fig:result_vs_FSI}. The data agrees well with generator and FSI model predictions, with $\chi^2$ values below nine, the number of differential bins, in all cases. Bins 5, 6, 8, and 9, which cover the kinematic region with highest purity and signal abundance, all prefer a higher signal cross section than that of all models. The data preference for a larger cross section is emphasized when studying the integrated cross section, summarized in Tab.~\ref{tab:integrated_cross_section_combined}. Among all the models investigated, GENIEv3 \texttt{AR23\_20i} and \texttt{G18\_10X} show the largest disagreement, both at a statistical tension with data equivalent to 1.7$\sigma$. The best agreement between data and simulation for the integrated cross section is found with \texttt{G18\_10d}.

\begin{table}[htpb!]
\centering
\caption{Integrated cross section values in the RoI with their associated $p-$values for different generators and FSI models.}
\label{tab:integrated_cross_section_combined}
\begin{tabular}{lrc}
\hline
Model & $\sigma$ (cm$^2$ / nucleon) & p-value \\
\hline
Data & $(6.07 \pm 1.22 )\times 10^{-41}$ & - \\
\hline
GENIEv3 AR23\_20i\_00\_000 & $4.02 \times 10^{-41}$ & 0.093 \\
NuWro v19.02.2 & $4.33 \times 10^{-41}$ & 0.153 \\
NEUT v5.6.2 & $4.11 \times 10^{-41}$ & 0.108 \\
\hline
G18\_10X & $3.99 \times 10^{-41}$ & 0.089 \\
G18\_10a & $4.07 \times 10^{-41}$ & 0.100 \\
G18\_10b & $4.24 \times 10^{-41}$ & 0.133 \\
G18\_10c & $4.42 \times 10^{-41}$ & 0.176 \\
G18\_10d & $5.26 \times 10^{-41}$ & 0.507 \\
\hline
\end{tabular}
\end{table}

{\em Conclusions.} Using the largest data sample of selected NC$1\pi^+$ interactions to date, the double-differential cross section for this channel has been presented for the first time. Overall, the differential cross-section results agree with tested model predictions, though we observe a statistically weak preference for a cross section approximately 30\% higher than most model expectations. The variations observed across FSI models are greater than those between neutrino interaction generators, and could significantly reduce the observed normalization discrepancy. This highlights the critical need to enhance our understanding of final-state interactions. We anticipate that NC$1\pi^+$ predictions will benefit from forthcoming high-statistics neutrino interaction measurements from T2K and other experiments. In particular, future measurements that precisely measure final-state hadrons in interactions both with and without final-state pions and utilizing transverse kinematic imbalance observables will be especially valuable to inform existing nuclear ground state and final-state-interaction models, directly benefiting NC$1\pi^+$ model predictions.

This work represents the first study of this interaction primarily with sub-GeV neutrinos, enabled by T2K's neutrino beamline, with dedicated emphasis on achieving a model-independent cross-section extraction, as discussed in Ref.~\cite{Abe:2025qki}. The results can be used to constrain NC$1\pi^\pm$ background events for T2K and the future Hyper-Kamiokande experiment, planned to start operations in 2027.

DUNE, using an Argon target and a higher neutrino energy, will likely benefit from additional studies on this channel. The selection strategy developed for this study, may guide future NC$1\pi^\pm$ measurements at other experiments.

This work establishes the foundation for more precise measurements of this channel in T2K and Hyper-Kamiokande using the ND280 detector. With the recent upgrade of the ND280 detector~\cite{T2K:2019bbb} and planned increases in data collection over the coming years, we expect continued improvements in the understanding of this channel, enabling more precise measurements of neutrino oscillation parameters. We anticipate that the upgraded ND280 geometry, with an enhanced angular acceptance, should allow to greatly extend the phase space of this measurement in the future to regions where the pion is emitted at large angles with respect to the beam, perhaps even including samples with backwards-going pions. An increase in the signal purity is also expected thanks to the better granularity and timing capabilities of the SuperFGD hydrocarbon target and the addition of Time-of-flight panels in the detector. Unlike the target used in this analysis, SuperFGD is also capable of identifying neutrons, opening new avenues to further understand this channel.

The measurement results have an associated data release that can be found in Ref.~\cite{T2K_release}.

\section*{Acknowledgments}
The T2K collaboration would like to thank the J-PARC staff for superb accelerator performance. We thank the CERN NA61/SHINE Collaboration for providing valuable particle production data. We acknowledge the support of MEXT,   JSPS KAKENHI  and bilateral programs, Japan; NSERC, the NRC, and CFI, Canada; the CEA and CNRS/IN2P3, France; the Deutsche Forschungsgemeinschaft (DFG 397763730, 517206441), Germany; the NKFIH  (NKFIH 137812 and TKP2021-NKTA-64), Hungary; the INFN, Italy; the Ministry of Science and Higher Education (2023/WK/04) and the National Science Centre (UMO-2018/30/E/ST2/00441, UMO-2022/46/E/ST2/00336 and UMO-2021/43/D/ST2/01504), Poland; the RSF (RSF 24-12-00271) and the Ministry of Science and Higher Education, Russia; MICINN, ERDF and European Union NextGenerationEU funds and CERCA program and Generalitat de Catalunya, Spain; the SNSF and SERI, Switzerland; the STFC and UKRI, UK; the DOE, USA; and NAFOSTED (103.99-2023.144,IZVSZ2.203433), Vietnam. We also thank CERN for the UA1/NOMAD magnet, DESY for the HERA-B magnet mover system, the BC DRI Group, Prairie DRI Group, ACENET, SciNet, and CalculQuebec consortia in the Digital Research Alliance of Canada, and GridPP in the United Kingdom, the CNRS/IN2P3 Computing Center in France and NERSC, USA. In addition, the participation of individual researchers and institutions has been further supported by funds from the ERC (FP7), “la Caixa” Foundation, the European Union’s Horizon 2020 Research and Innovation Programme under the Marie Sklodowska-Curie grant; the JSPS, Japan; the Royal Society, UK; French ANR and Sorbonne Université Emergences programmes; the VAST-JSPS (No. QTJP01.02/20-22);  and the DOE Early Career programme, USA. For the purposes of open access, the authors have applied a Creative Commons Attribution license to any Author Accepted Manuscript version arising.

\bibliographystyle{apsrev4-1}
\bibliography{biblio}

\end{document}